# Bright X-ray bursts from 1E1724-3045 in Terzan 2

M. Cocchi[(1)], A. Bazzano[(1)], L. Natalucci[(1)] and P. Ubertini[(1)]
J. Heise[(1)], E. Kuulkers[(1)] and J. J. M. in 't Zand[(1)]

[(1)]*Istituto di Astrofisica Spaziale), via Fosso del Cavaliere, 00133 Roma, Italy*
[(2)]*Space Research Organisation Netherlands (SRON), Sorbonnelaan 2, 3584 CA, Utrecht, the Netherlands*

**ABSTRACT**. During about 3 years wide field monitoring of the Galactic Centre region by the WFC telescopes on board the *BeppoSAX* satellite, a total of 14 type-I X-ray bursts were detected from the burster 1E 1724-3045 located in the globular cluster Terzan 2. All the observed events showed evidence of photospheric radius expansion due to Eddington-limit burst luminosity, thus leading to an estimate of the source distance (~7.2 kpc). Preliminary results of the analysis of the bursts are presented.

## INTRODUCTION

Since its discovery [14], the globular cluster source 1E 1724-3045, has been repeatedly studied by several satellite experiments both in the "classical" (~1-20 keV) X-ray energy band (e.g. *EINSTEIN, EXOSAT, TTM, ROSAT, BeppoSAX, RXTE, ASCA*) and in the hard X-rays (*GRANAT, BeppoSAX, RXTE*).

The source is persistently bright, though variable, and is one of the few X-ray bursters showing persistent emission up to the soft Gamma-rays [1],[17],[16].

Type-I X-ray bursts from 1E 1724-3045 were observed since the very first observations performed with OSO-8 [14]. In particular, *EINSTEIN* detected a bright burst showing photospheric radius expansion [7],[15], allowing an estimate of the source distance (~7 kpc) which is consistent with what is obtained taking into account the measured reddening of the globular cluster Terzan 2 [13].

Bursting activity identifies the source as a weakly magnetized neutron star in a low mass binary (*LMXB*) system.

Soft X-ray observations have shown that the persistent spectrum is best-fitted by a power law of photon index ~2.0-2.4 [9],[12],[18]. A higher spectral index (~3.0) is obtained in the hard X-rays by *SIGMA* [5],[6]. More recently, simultaneous *SAX* and *RXTE* observations demonstrated that the wide band (1-200 keV) spectrum of 1E 1724-3045 is actually power law below ~30 keV, attenuated at high energies by an exponential cutoff at ~70 keV [8],[2]. This is interpreted as the result of the Comptonization of soft photons in a spherical scattering region of electron temperature ~30 keV and optical depth ~3. Besides that, the *RXTE* and *SAX* observations suggested the presence of an additional soft component, most likely a multicolor disk blackbody, as confirmed by recent *ASCA* results [3].

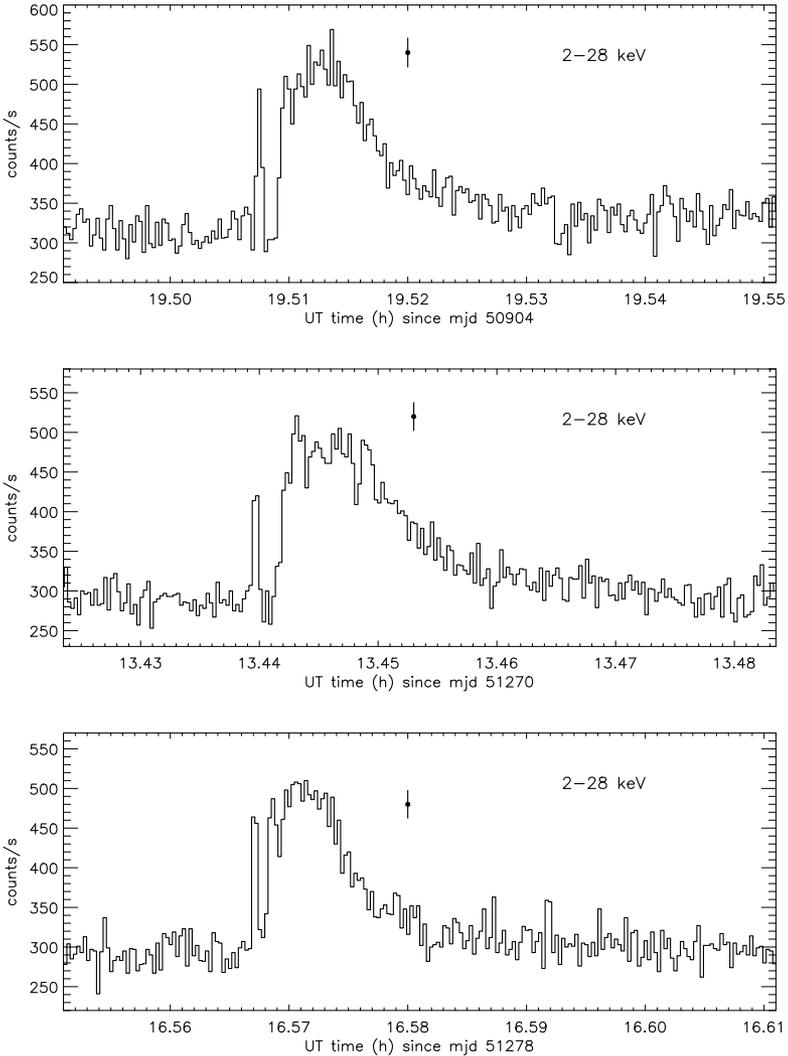

**FIGURE 1.** 2-28 keV time history of three different bursts of 1E 1724-3045.

## OBSERVATION AND DATA ANALYSIS

The *Wide Field Cameras (WFC)* on board *BeppoSAX* satellite consist of two identical coded mask telescopes [10]. The two cameras point at opposite directions each covering 40°×40° field of view. With their source location accuracy in the range 1'-3', a time resolution of 0.244 ms, and an energy resolution of 18% at 6 keV, the *WFC*s are very effective in studying hard X-ray (2-28 keV) transient phenomena.

The imaging capability, combined with the good instrument sensitivity (5-10 mCrab in $10^4$s), allows accurate monitoring of complex sky regions like the Galactic Bulge.

The data of the two cameras is systematically searched for bursts or flares by analyzing the time profiles of the detector in the 2-11 keV energy range with a 1s time resolution. Reconstructed sky images are generated in coincidence with any statistically meaningful enhancement, to identify possible bursters. The accuracy of the reconstructed position, which of course depends on the intensity of the burst, is typically better than 5'. This analysis procedure allowed for the identification of ~700 X-ray bursts in a total of about 2 Ms WFC net observing time (see e.g. [4]).

Among the detected bursts, a total of 14 was observed from a sky position consistent with that of Terzan 2.

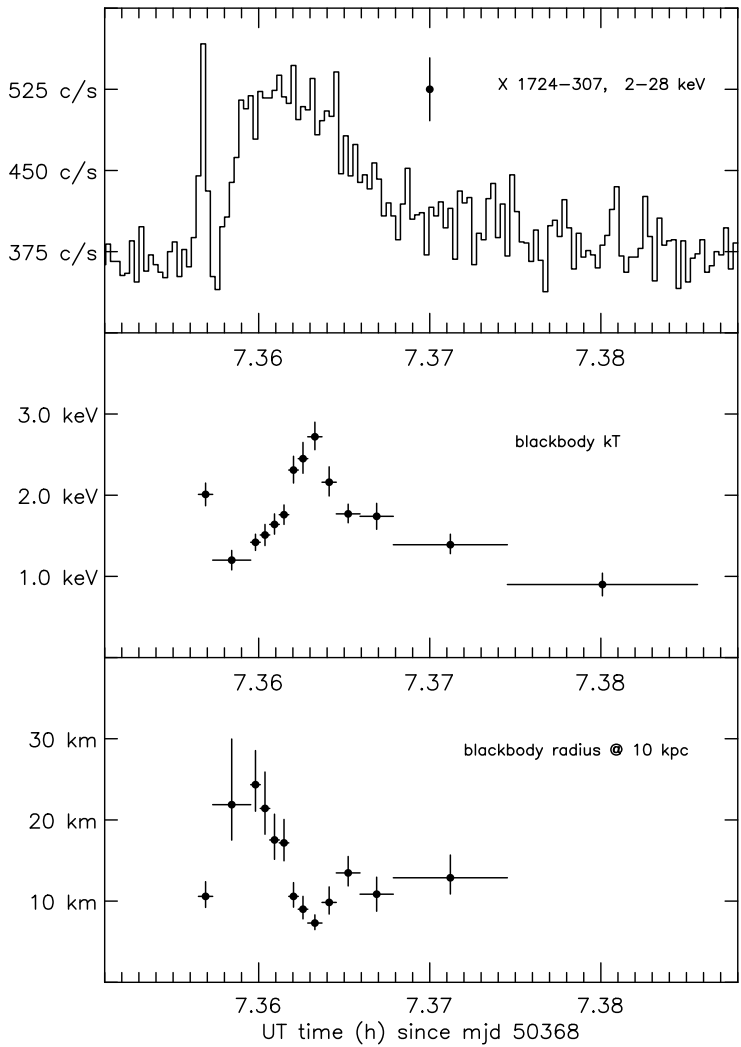

**FIGURE 2.** Results of time resolved spectroscopy on the mjd 50368 burst. Upper panel: 2-28 keV time history of the burst; Middle panel: time history of the measured blackbody temperature; Lower panel: time history of the blackbody radius, assuming a 10 kpc distance.

**TABLE 1.** Summary of the results of the time resolved spectral analysis of the mjd 50368 burst (assumed $N_H = 10^{22}$ cm$^{-2}$).

| Data set | Time range (T0= 7.35648 UT) | blackbody kT (keV) | $\chi^2 R$ (27 d.o.f.) | R (km) @10 kpc |
|---|---|---|---|---|
| Whole burst | T0 – T0+68 s | 1.80 ± 0.04 | 1.14 | 11.2 ± 0.5 |
| A | T0 – T0+3 s | 2.01 ± 0.14 | 0.65 | 10.6 +1.8 / -1.4 |
| B | T0+3 s – T0+11 s | 1.20 ± 0.12 | 1.20 | 21.9 +8.1 / -4.4 |
| C | T0+11 s – T0+13 s | 1.42 ± 0.10 | 0.73 | 24.3 +4.2 / - 3.3 |
| D | T0+13 s – T0+15 s | 1.51 ± 0.13 | 1.11 | 21.4 +4.5 /- 3.2 |
| E | T0+15 s – T0+17 s | 1.64 ± 0.12 | 1.08 | 17.5 +3.2 / -2.4 |
| F | T0 +17s – T0+19 s | 1.76 ± 0.12 | 1.68 | 17.2 +2.9 / -2.2 |
| G | T0+19 s – T0+21 s | 2.31 ± 0.16 | 1.01 | 10.6 +1.7 / -1.3 |
| H | T0+21 s – T0+23 s | 2.45 ± 0.19 | 0.94 | 9.0 +1.6 / -1.2 |
| I | T0+23 s – T0+26 s | 2.72 ± 0.17 | 0.96 | 7.3 +1.0 / -0.8 |
| J | T0+26 s – T0+29 s | 2.16 ± 0.18 | 0.91 | 9.8 +1.9 / -1.4 |
| K | T0+29 s – T0+34 s | 1.77 ± 0.11 | 1.36 | 13.5 +2.0 / -1.6 |
| L | T0+34 s – T0+41 s | 1.74 ± 0.16 | 0.75 | 10.9 +2.1 / -2.1 |
| M | T0+41 s – T0+65 s | 1.39 ± 0.12 | 0.65 | 12.9 +2.8 / -2.0 |
| N | T0+65 s – T0+95 s | 0.90 ± 0.14 | 1.09 | 22.6 +8.1 / -8.1 |

A preliminary analysis of the burst data showed evidence of photospheric radius expansion on all the detected bursts. In particular, the so-called *precursor event* is always observed in the time histories of the bursts. The precursors last in average 1-2 s, and are followed a few seconds later (~5 s in average) by the main burst. In Figure 1 and Figure 2 a sample of 4 bursts is shown; the time profiles are obtained in the full WFC energy band (2-28 keV).

Detailed time resolved spectroscopy of the mjd 50368 burst was performed (Figure 2, Table 1), in order to study the time evolution of relevant burst parameters. To better constrain the fits, the $N_H$ parameter was kept fixed to $10^{22}$ cm$^{-2}$. The burst spectra are all consistent with an absorbed blackbody model, typical of type-I X-ray bursts, originating by helium thermonuclear flashes onto the surface of a neutron star (see [11] for a review). The time history of the blackbody temperature also points to type-I bursts, since the observed spectral softening is commonly regarded as the result of the cooling of the neutron star photosphere after the flash.

The time history of the blackbody radius is consistent with an adiabatic expansion of the photosphere, most likely due to Eddington luminosity of the burst; the data is consistent with a radius expansion of a factor ~2.5 just after the precursor event. After the subsequent contraction of the emitting region, the burst behaves as a more typical type-I, showing the color softening and an almost constant blackbody radius. The burst decay is exponential with a characteristic time of 24.1 ± 1.7 s.

Eddington-luminosity X-ray bursts allow for an estimate of the source distance. The 2-28 keV peak intensities of the 4 analyzed bursts are all consistent with a weighted average of 1.01 ± 0.03 Crab, which extrapolates to a bolometric intensity of (3.23 ± 0.10) ×10$^{-8}$ erg cm$^{-2}$ s$^{-1}$. The consistency of the peak intensities of all the analyzed bursts with a constant value supports the adoption of the peak bolometric luminosities of super-Eddington bursts as a standard candle. Assuming a $2\times10^{38}$ erg s$^{-1}$ Eddington luminosity for a 1.4 M$_o$ neutron star and using standar burst parameters, we obtain for 1E 1724-307 a distance value d = 7.2 ± 0.2 kpc in agreement with previous results [15].

From the average blackbody radius of the cooling track and for the estimated distance of 7.2 kpc we derive an average radius of ~7 km for the blackbody emitting region during the burst. On the other hand, if we assume the average luminosity of Eddington-limit bursts proposed by Lewin, van Paradijs and Taam [11], i.e. (3.0 ± 0.6) $\times10^{38}$ erg s$^{-1}$, we obtain a slightly larger distance, d = 8.8 ± 1.0 kpc.